\documentclass[twocolumn,preprintnumbers,amsmath,amssymb]{revtex4-1}

\usepackage{amssymb}
\usepackage{amsbsy}
\usepackage{amsmath}
\usepackage[dvips]{graphics,epsfig}

\begin{document}

\title{Calculation of the melting point of alkali halides by means of computer simulations}

\author{J.L. Aragones, E. Sanz, C. Valeriani and C. Vega \footnote{cvega@quim.ucm.es}}
\affiliation{Departamento de Qu\'{\i}mica F\'{\i}sica, Facultad de Ciencias Qu\'{\i}micas,
Universidad Complutense de Madrid, 28040, Spain.}

\newcommand{\prom}[2]{\left \langle #1 \right \rangle_{#2}}
\newcommand{\du}{\mbox{$\,$} \mathrm{d}}
\newcommand{\Ppar}{p_{\parallel}}
\newcommand{\Pper}{p_{\perp}}
\newcommand{\derpar}[3]{
        \left(\frac{\partial #1}{\partial #2}\right)_{#3}
                      }
\newcommand{\Eq}[1]{Eq.~(\ref{eq:#1})}
\newcommand{\rvec}[2]{\mathbf{r}_{#1}^{#2}}
\newcommand{\qvec}[2]{\mathbf{q}_{#1}^{#2}}
\newcommand{\tvec}[2]{\mathbf{t}_{#1}^{#2}}

\date{\today}

\begin{abstract}

In this manuscript we study the liquid-solid coexistence of NaCl-type alkali halides, 
described by interaction potentials such as 
Tosi-Fumi (TF), Smith-Dang (SD) and Joung-Cheatham (JC), 
and compute their melting temperature  (T$_m$) at 1 bar 
via three independent routes: 1) liquid/solid direct coexistence, 2) free-energy calculations and 
3) Hamiltonian Gibbs-Duhem integration. 
The melting points obtained by the three routes
are consistent with each other.
The calculated T$_m$ of the Tosi-Fumi model of NaCl is in good agreement with the experimental value as well as 
with other numerical calculations. However, the other two models considered  for
NaCl, SD and JC, overestimate the melting temperature of NaCl by more than 200~K.
We have also computed the melting temperature of other alkali halides using the Tosi-Fumi
interaction potential and observed that the predictions are not always as close to the
experimental values as they are for NaCl. 
It seems that there is still room for improvement 
in the area of force-fields for alkaline halides, given  that so far  most  models are still unable to 
describe a simple yet important property such as the melting point.

\end{abstract}

\maketitle

\twocolumngrid

\section { Introduction }

Alkali halides are inorganic compounds composed of an alkali metal and a halogen. 
The most abundant by far in Nature is Sodium Chloride (NaCl). NaCl in its solid form has a cubic structure
(usually denoted as NaCl structure) and melts at a relatively high temperature (around 1070~K at ambient pressure)
as a consequence of its high lattice energy.
NaCl dissolves in polar solvents, such as water, to give ionic solutions that 
 contain highly solvated anions and cations, relevant for the functioning of biological organisms.

In the last century, there have been many thorough studies aimed at quantitatively describing the  physical properties 
of alkali halides.
To a first approximation alkali halides are two-component mixtures of atomic
anions and cations which interact through a spherically symmetrical
potential. The two unusual features of ionic crystals are the slow decay of
the interaction potential, which causes  long-range structural
correlations, and the strength of the attractive cation-anion interactions.
The simplest model for a molten salt is the restricted primitive model (RPM) , in
which ions are modelled by hard spheres with positive and negative unit 
charges. The RPM has been studied by a number of 
groups\cite{PRE_57_6944_1998,JCP_1999_110_01581,PRE_2004_70_016114,hynninen,PRL_90_135506_2003,pre_madrid,JCP_2003_119_00964}. 
Back in 1919,  Born developed a numerical model to estimate 
the energy of an ionic crystal~\cite{PG_21_13_1919}. Pauling, 
based on Born's primitive model, studied the effect of the ions' sizes on ionic salts~\cite{ZfK_67_377_1928,JACS_50_1036_1928} 
and  Mayer evaluated the role of polarizability and dispersive forces on alkali halides~\cite{JCP_1933_1_270} 
and proposed, together with Huggins, a generalization of Born's  repulsive energy~\cite{JCP_1933_1_643}. 
More recently, several ion-ion interaction potentials have been developed in order to numerically simulate alkali halides. 
In 1962 Tosi and Fumi, fitting the Huggins-Mayer dispersive energy to crystallographic data, 
proposed an empirical potential parameterizing the repulsive part of the NaCl alkali halide interactions~\cite{JPCS_1964_25_00031}
and including the dipole-quadrupole van der Waals term, that imitated the polarization
distortion of the electronic  cloud.
The advantages and disadvantages of the Tosi-Fumi (TF) potential have been later underlined by Lewis and coworkers~\cite{FT_71_301_1975}.
The model in general predicts good densities and lattice energies of alkaline halides. However it is unable to reproduce 
dynamical features of ionic crystals such as the
phonon dispersion curves. Also the model predicts incorrectly the similarity of like pair distribution functions $g^{--}$ and $g^{++}$ which is absent in the experiments \cite{JPCM_5_2687_1992}.
The TF model was used for the first time in a numerical simulation by Adams and McDonald~\cite{JPCS_1974_7_2761}, who  
obtained a remarkably good agreement between numerical and experimental results.
In recent years, the Tosi-Fumi potential has also been used to study liquid/solid phase transitions in alkali halides: 
Valeriani {\em et al.} studied homogeneous  crystal nucleation in molten NaCl~\cite{JCP_122_194501_2005} and 
Chen and Zhu carried out homogeneous nucleation studies of other alkali halides, such as KBr~\cite{CJIC_20_1050_2004}
and NaBr~\cite{JMST_680_137_2004}.
Zykova-Timan {\em et al.} studied packing issues related to the interfacial free-energy
between the liquid and solid phases~\cite{SS_566_794_2004,PRL_100_036103_2008}.
In all these works knowledge of the melting temperature $T_{m}$ was needed. 
In fact, the melting point of NaCl has been estimated for the TF model by several groups.
In 2003, Anwar {\em et al.} determined by free-energy calculations
the melting temperature of NaCl for the Tosi-Fumi model~\cite{JCP_2003_118_00728}. Later on similar results were
obtained by Eike {\em et al.}~\cite{brennecke} and Mastny {\em et al.}~\cite{JCP_122_124109_2005} and
by  Zykova-Timan {\em et al.} \cite{zykova3}. 
Molecular simulations have also been performed to 
analyze experimental results~\cite{PRL_78_4589_1997,PRB_53_556_1996}.
In some  works the melting temperature was calculated by direct heating of the solid, hence the
thermal instability limit rather that the melting temperature was calculated~\cite{PRB_61_11928_2000}.
More recently in order to compute T$_m$ Belonoshko {\em et al.} performed two-phase simulations of NaCl and LiF 
~\cite{AM_81_303_1996,PRB_61_11928_2000}.

Besides TF, another popular model used to describe NaCl was proposed more recently by 
Smith and Dang (SD)~\cite{JCP_1994_100_03757,JCP_1993_99_6950} who presented an interaction potential
where the ion-ion interactions were Lennard-Jones like. This model potential 
has become quite popular when studying NaCl in water solutions, even though the properties of its solid
phase are still unknown.
Similar to the SD potential, Joung-Cheatham (JC) presented another interaction potential 
for NaCl where the ion-ion interactions were Lennard-Jones like and proposed 
several NaCl force-fields tailored to be used in a water solution~\cite{JPCB_112_9020_2008}. 
Somewhat surprisingly the melting point of the SD and JC/NaCl potentials are still unknown.

Good model potentials for  alkali halides are also useful  to study ionic solutions. 
Simulations of alkali halides dissolved in water have proved useful to study thermodynamic mixing
properties, such as the cryoscopic descent of the melting temperature \cite{jungwirth07}.
The properties of alkali halides solutions have been studied at low temperatures
in order to localize the hypothesised second critical point of water \cite{corradini10,ISI:000287066600016,N_1992_360_00324_nolotengo}.
The solubility of NaCl in water has also received certain interest \cite{JPCB_115_7849_2011,sanz_nacl,aragones_nacl,JPCB_113_13279_2009,JPCB_109_12956_2005,paluch,lisal_2012}.
However we should underline  that in order to determine the solubility of a salt, the chemical potential of the solid should be known.

Thus, it is relevant to quantitatively compare several interaction potentials in order to estimate their 
efficiency in mimicking the properties of  NaCl-type alkali halides.
To this aim, in this manuscript
we evaluate the liquid-solid equilibrium  of such systems. We have computed
the melting temperature (at normal pressure) for three alkali halides, emphasizing NaCl, 
for three interaction models: the Born-Mayer-Huggins-Tosi-Fumi  (TF)~\cite{JCP_1933_1_270,JCP_1933_1_643,JPCS_1964_25_00031, JPCS_1964_25_00045}, 
the Smith-Dang  (SD)~\cite{JCP_1994_100_03757} and 
the Joung-Cheatham (JC) potential~\cite{JPCB_112_9020_2008}.
The three potentials are two-body and non-polarizable model potentials, characterized by 
a repulsive term, a short-range attraction and a long-range Coulombic interaction term.
To calculate the melting temperature (T$_m$) we used three
independent routes: 1) liquid/solid  direct coexistence 2) free-energy calculations and
3) Hamiltonian Gibbs-Duhem integration. We have found that the value obtained for 
the T$_m$ of the TF/NaCl is in good agreement with the experiment, in contrast with the results obtained for
the SD/NaCl and JC/NaCl models. Therefore, we conclude
that the TF/NaCl model is the most suitable for studies of pure NaCl.
Using Hamiltonian Gibbs-Duhem integration we have also evaluated T$_m$ for other TF/alkali halides and concluded 
that the quality of the results obtained with the TF potential depends on the alkali halide chosen.
The TF model provides good predictions for alkali halides that involve K$^+$, Cl$^-$, Na$^+$,
Br$^-$ and Li$^+$ ions, whereas it performs much worse for alkali halides involving Rb$^+$ or F$^-$ ions.

The manuscript is organized as follows: we first introduce the interaction potentials of the 
alkali halides under study, i.e. Born-Mayer-Huggins-Tosi-Fumi potential (TF), 
the Smith-Dang potential (SD) and the Joung-Cheatham (JC) potential. 
Then we describe the three simulation routes followed to compute
their melting temperature: 1) the liquid/solid  direct coexistence
 2) the Einstein crystal/molecule for the solid and the thermodynamic integration for the liquid and 
3) the Hamiltonian Gibbs-Duhem integration.
To conclude we will present our results obtained for different alkali halides.

\section{Simulation methods} \label{theory}

The  interaction potentials we used are two-body and non-polarizable model potentials, 
each of them characterized by a repulsive term, a short-range attractive and a long-range Coulomb interaction term.
The TF  model potential has the following form: 
\begin{equation}
\label{TF}
U(r_{ij})=A_{ij}e^{-r_{ij}/\rho_{ij}}-\frac{C_{ij}}{r_{ij}^{6}}-\frac{D_{ij}}{r_{ij}^{8}}+\frac{q_i q_j}{4 \pi \epsilon_o r_{ij}}
\end{equation}
where $r_{ij}$ is the distance between two ions with charge $q_{i,j}$, the first term is the Born-Mayer repulsive
 term, $-\frac{C_{ij}}{r_{ij}^{6}}$ and $-\frac{D_{ij}}{r_{ij}^{8}}$ are the Van der Waals attractive interaction
terms and the last term corresponds to  the Coulomb interaction.
The parameters $A_{ij}$, $\rho_{ij}$, $C_{ij}$ and  $D_{ij}$ for the TF/alkali halides 
are given as Supplementary Material\cite{material_suplementario}. 

Both the SD and the JC model potentials  can be written using the following expression:
\begin{equation}
U(r_{ij}) = 4 \epsilon \left[ \left( \frac{\sigma_{ij}}{r_{ij}} \right)^{12} - \left( \frac{\sigma_{ij}}{r_{ij}}\right)^6 \right] +\frac{q_i q_j}{4 \pi \epsilon_o r_{ij}}
\label{SDJC}
\end{equation}
where $r_{ij}$ is the distance between two ions with charge $q_{i,j}$. The first term is 
Lennard-Jones-like, and its parameters are given in the Supplementary Information file\cite{material_suplementario}. 
The last term in Eq.~\ref{SDJC} corresponds to the Coulomb interaction.
For the JC potential, we are going to use the parameters introduced to simulate NaCl in SPC/E water~\cite{JPCB_112_9020_2008}.

For the SD and JC the crossed interaction parameters are obtained using the 
Lorentz-Berthelot combining rules \cite{lorentz,berthelot}.
It is interesting to point out that the TF model was developed to study ionic crystals and pure
alkali halides in the solid phase whereas the SD was obtained to model Na$^+$ and Cl$^-$
in water. The JC was fitted to model NaCl both in the solid phase and in aqueous solutions.
 In what follows, we shall refer to ions as particles.

When computing T$_m$ for the TF/NaCl interaction potential, we follow three independent routes:
1) liquid/solid direct coexistence, 2) free-energy calculations and 
3)Hamiltonian Gibbs-Duhem integration.
For the SD/NaCl and JC/NaCl and for other TF/alkali halides
we use the first and the third route.

\subsection{Route 1. Liquid/solid direct coexistence}

The first route we follow to compute the melting temperature is by means of the 
liquid/solid direct coexistence, originally proposed in Ref.~\cite{ladd77,ladd78,morris02,morris04,karim88}.
To start with, we generate an equilibrated configuration of the NaCl solid phase in contact with its liquid. 
After having prepared the liquid-solid configuration, we run several NpT simulations (with anisotropic scaling,
so that each side of the simulation box changes independtly)   
at different temperatures and always at a pressure of 1 bar.  
Direct coexistence simulations should be performed in the Np$_z$T ensemble,
where z is the axis perpendicular to the fluid-solid interface and $x$ and $y$ have been chosen 
carefully to avoid the presence of stress. However, we have recently shown that NpT simulations
(with anisotropic scaling) provide proper results for sufficiently large systems \cite{noya_hs_interface,ramon06}. 
Depending on the temperature, the system evolves towards complete freezing or melting of the sample.  
Since we do not know the location of the melting temperature, we simulate the system in a wide range
of temperatures and identify the melting temperature $T_m$ as the average of the highest temperature
at which the liquid freezes and the lowest temperature at which the solid melts. A more elaborate procedure
would require to estimate the slope of the growth rate of the solid-fluid interface ( which is positive at 
temperatures below the melting point and negative at temperatures above the melting point) and to locate 
the temperature at which the growth rate of interface is zero (i.e the melting point). Since the growth of
the solid  
is a stochastic process the determination of the growth rate requires to accumulate statistics over many different trajectories making
the calculations quite expensive\cite{kusalik_ice_growth,alemanes}.  
The procedure used here is simpler and has provided reliable results for other systems as 
hard spheres\cite{noya_hs_interface,ramon06}
or water\cite{ramon06} although admittedly it yields a somewhat
larger error in the estimate of the melting point temperature obtained from direct coexistence simulations (of about 5K). 

\subsection{Route 2. Free-energy calculations}

In this route, we first compute the Helmholtz free-energy of both the solid and the liquid phase; 
next, we estimate the Gibbs free-energy ($G$) by simply adding $pV$.
To compute the melting temperature we perform thermodynamic integration as a function of temperature 
at constant pressure to evaluate where the chemical potentials ($\mu = G/N$) of both phases coincides.
To estimate the free-energy of the bulk solid phase we 
use the Einstein crystal~\cite{frenkel84} and the Einstein molecule methods~\cite{vega_noya,JCP_129_2008}.
Both methods are based on the calculation of the free-energy difference between the target solid and a reference 
system at the solid equilibrium-density for the given thermodynamic conditions
(obtained with an NpT simulation).
The reference system of the Einstein crystal method consists of an ideal solid whose free-energy
can be analytically computed (an ``Einstein crystal'' with the center of mass fixed, where the
inter-molecular interactions are neglected and particles are bound to their lattice equilibrium positions by a harmonic 
potential with strength $\Lambda_E$). 
The Einstein molecule method differs from the previous one due to the fact that we now fix only the position
of one  particle instead of the center of mass. 

Thermodynamic integration is performed in two steps~\cite{vega_review}: 
1) we evaluate the free-energy difference ($\Delta$ A$_1$) between the ideal Einstein crystal
and the Einstein crystal in which particles interact through the Hamiltonian of the original solid (``interacting'' Einstein crystal). 
2) Next we calculate the free-energy difference  ($\Delta$ A$_2$) between  the 
interacting Einstein crystal and the original solid, by means of thermodynamic integration: 
$U(\lambda)=\lambda U_{sol}+( 1 - \lambda) (U_{Ein-id}+U_{sol})$), where  $U_{Ein-id}$ represents the energy of the interacting Einstein crystal, 
$U_{sol}$ the one of the original solid, and $\lambda$ is the coupling parameter that allows us to integrate from the 
interacting Einstein crystal ($\lambda=0$)  to the desired solid ($\lambda=1$).
The final expression of the Helmholtz free-energy  $A^{NaCl}_{sol}(T,V)$ coming from the Einstein crystal/molecule calculations is \cite{vega_review}:
\begin{equation}
    A_{sol}^{NaCl}(T,V) =  A_0(T,V) +  \Delta A_1(T,V)  +  \Delta A_2(T,V)  
\label{einstein}
\end{equation}
where A$_0$ is the free energy of the reference system, whose  analytical expression is slightly
different in the Einstein crystal and Einstein molecule (see Ref.~\cite{vega_review}).
$\Delta$A$_1$ and  $\Delta$A$_2$ are computed in the same way in the  Einstein
crystal and Einstein molecule methods
(the only difference being the choice of the point that remains fixed in the simulations,
whether the system's center of mass or a reference particle's center of mass).  
It has been shown that, since the free energy of a solid is uniquely defined,
its value does not depend on the method used to compute it and the two methodologies give exactly the same results~\cite{vega_review}.
It is convenient to set the thermal De Broglie wave length of all species 
to 1~\AA, and the internal partition functions of all species to one.
These arbitrary choices affect the value of the free energy but does not affect
phase equilibria (provided the same choice is adopted in all phases).

To estimate the free-energy of the bulk liquid we use Hamiltonian thermodynamic integration 
as in Ref.~\cite{JCP_2003_118_00728}, 
calculating the free-energy difference between the liquid alkali halide and a reference Lennard-Jones (LJ) liquid for which the 
free-energy is known. 
Starting from an equilibrated NaCl liquid, we perturb 
the Hamiltonian of the system so that  each ion is gradually transformed into 
a LJ atom. The path connecting both states is given by:
\begin{equation}
\label{HGDI}
U(\lambda) = \lambda U^{LJ} + (1-\lambda) U^{NaCl}
\end{equation}
where $U^{LJ} $/$U^{NaCl}$ are the total energies of the Lennard Jones and NaCl fluids, respectively, and 
$\lambda$ is the coupling parameter ($\lambda=0$ corresponds to NaCl whereas $\lambda$=1 to a LJ fluid). 
The Helmholtz free-energy of a NaCl (A$^{NaCl}_{liq}$) is computed as:
\begin{eqnarray}
\label{freeliq}
A_{liq}^{LJ}(T,V) & = & A_{liq}^{NaCl}(T,V) + \int_{\lambda=0}^{\lambda=1} \langle U^{LJ} - U^{NaCl} \rangle_{N,V,T,\lambda} d\lambda \nonumber \\
& = & A_{liq}^{NaCl}(T,V) + \Delta A_{liq}^{LJ}(T,V)
\end{eqnarray}
Given that the Lennard-Jones free-energy (A$_{liq}^{LJ}$) is already known \cite{FPE_100_1994} and the 
integral in the Eq.~\ref{freeliq} ($\Delta A_{liq}^{LJ}$) can be numerically evaluated, 
we can determine the free-energy A$^{NaCl}_{liq}$ of the liquid alkali halide.
In order to estimate the integral in Eq.~\ref{freeliq}, we choose 20 values of
$\lambda$ between 0 and 1, 10 of them 
equally spaced from 0.000 to 0.95 and the remaining 10  from 0.95 to 1.000, and 
integrate each region using the Simpson integration method.
This choice originated from then fact that when $\lambda$ has a value
close to 1.0, the integrand changes abruptly, dropping to the dispersive energy of a LJ.
The Lennard-Jones free-energy consists of two terms: 
$A_{liq}^{LJ}(T,V) = A_{liq}^{LJ,id}(T,V) + A_{liq}^{LJ,res}(T,V)$, where A$_{liq}^{LJ,res}(T,V)$ is the excess 
and A$_{liq}^{LJ,id}(T,V)$  the ideal part. A$_{liq}^{LJ,res}$ for a Lennard-Jones fluid has been already
computed for  a broad range of temperatures and densities in Ref.~\cite{MP_1993_78_0591_photocopy,FPE_100_1994}. 
The free energy of the ideal gas of a mixture of N$_{Na}$ and N$_{Cl}$ is given by 
\begin{eqnarray}
\frac{1}{k_B T} A_{liq}^{LJ,id}  (T,V) & = & N_{Na} ln(\rho_{Na} \Lambda^{3}_{{Na}}) - N_{Na} \nonumber \\ 
& + & N_{Cl} ln(\rho_{Cl} \Lambda^{3}_{Cl}) - N_{Cl} \nonumber \\ 
& = & N[ln(\rho/2) - 1]
\end{eqnarray}
where N = N$_{Na}$+ N$_{Cl}$ is the total number of particles in the system with density
$\rho=\frac{N}{V}$ (and $\rho_{Na}=\frac{N_{Na}}{V}$ = $\rho_{Cl}=\frac{N_{Cl}}{V}$ = $\rho$/2)
and $\Lambda_i$ is the De Broglie thermal length ($\Lambda$$_i$ = h / $\sqrt{2 \pi m_i k_B T}$),
that we  arbitrarily set to 1~\AA, to be consistent with our choice for the solid phase.

 \subsection{Route 3. Hamiltonian Gibbs-Duhem integration}
 \label{HGDIsec}

The third route we follow to compute the melting temperature is by means of  
Hamiltonian Gibbs-Duhem thermodynamic integration as in Ref.~\cite{gdfilosofico1,gdfilosofico2,vegafilosofico}.
Starting from the liquid-solid coexistence point of a reference system ($A$), 
the Hamiltonian Gibbs-Duhem thermodynamic integration allows one to compute the coexistence point
of the system of interest (whose Hamiltonian is $B$) by resolving a Clapeyron-like differential equation.
In more detail, the Hamiltonian of the initial system (with energy $U_A$), whose two-phases coexistence
point is known, is connected to the one of the final system of interest (with energy $U_B$),
via the following expression:
\begin{equation}
\label{coupling}
U(\lambda)=\lambda U_B + (1-\lambda)U_A
\end{equation}
where $\lambda$ is the coupling parameter.  When two phases coexist (labeled as I and II):
$\mu_I(T,p,\lambda)=\mu_{II}(T,p,\lambda)$, 
being $\mu=G/N$ the Gibbs free-energy per particle 
(the chemical potential of each phase). Therefore, differentiating $\mu$ in both phases, 
we can write generalized Clapeyron equations for two coexisting phases as
\begin{eqnarray}\label{genclap}
v_I (T,p) dp -s_I (T,p) dT + \left( \frac{\partial \mu_I (T,p,\lambda)}{\partial \lambda} \right) d\lambda  = \nonumber \\ 
 =  v_{II}(T,p)  dp-s_{II} (T,p) dT + \left( \frac{\partial \mu_{II} (T,p,\lambda) }{\partial \lambda} \right) d\lambda
\end{eqnarray}
where $v$ and $s$ are the volume and entropy  per particle. If $\lambda$ is constant  we recover the well known Clapeyron equation. 
If the pressure is constant we obtain the slope of the coexistence line in the $\lambda$-T plane:
\begin{eqnarray}
\label{ham_GD_int}
\frac{dT}{d\lambda}=\frac{T[(\partial \mu_{II}/\partial \lambda)-(\partial \mu_{I}/\partial \lambda)]}{h_{II}-h_{I}}  =  \nonumber \\ 
= \frac{T[(\partial \mu_{II}/\partial \lambda)-(\partial \mu_{I}/\partial \lambda)]}{\Delta h}
\end{eqnarray}
knowing that, at coexistence
\begin{equation}
\label{meltentropy}
s_{II}-s_I = \frac{h_{II}-h_I}{T}. 
\end{equation}
When a liquid coexists with a solid, ($s_{II}-s_I$) is the melting entropy difference $\Delta s_m$, 
that can be easily computed as $\frac{\Delta h_m}{T_m}$. 
$\Delta$h is obtained from the NpT simulations at (p,$\lambda$,T) constants, whereas 
$\frac{\partial \mu}{\partial \lambda}=\langle \frac{\partial u(\lambda)}{\partial \lambda} \rangle_{N,p,T,\lambda}$, 
computed with an NpT simulation at different values of $\lambda$ in each phase.

Therefore, using Eq.~\ref{coupling}, 
the generalized Clapeyron equation can be written as:
\begin{equation}
\label{ham_GD_int2}
\frac{dT}{d\lambda}=\frac{T[\langle u_B-u_A\rangle^{II}_{N,p,T,\lambda}-\langle u_B-u_A\rangle^{I}_{N,p,T,\lambda}]}{\Delta h}
\end{equation}
where $u_B/u_A$ is the internal energy per particle when the interaction between particles is described by $U_B/U_A$. 
The numerical integration of the generalized Clapeyron equation in Eq.~\ref{ham_GD_int2}
yields the change of the coexistence temperature
(at constant pressure) due to the change in the Hamiltonian of the system, 
starting from the initial coexistence point (where interactions are 
described by $U_A$) to the final coexistence point (where interactions are described by $U_B$).

\section{Simulation Details}

When using the liquid/solid direct coexistence route to compute the melting temperature  
 we choose  systems containing either 1024 [512 solid/512 liquid], 2000 [1000solid/1000liquid]  or 4000 ions 
[2000 solid/2000 liquid].
When using the free-energy calculations route
to compute the melting temperature, we simulate systems of 1000 ions  
(the unit cell of NaCl contains 8 particles). 
For the Hamiltonian Gibbs-Duhem integration route we simulate systems of 1000 ions.

We simulated NaCl using the TF, the SD and the JC interaction potentials.
For the other alkali halides considered in this work we used only the TF potential. 
In this work  we truncated the non-Coulomb part of the potential at r$_c$=14~\AA~ and added tail corrections.
We used Ewald sums to deal with Coulombic interactions, truncating the real 
part of the Ewald sums at the same cut-off as the non-Coulombic  interactions and chose the
parameters of the Fourier part of the Ewald sums so that $\alpha \cdot r_c = 2.98$~\cite{frenkelbook,allen_book}.  
To calculate T$_m$ for the SD and JC models via direct coexistence we performed NpT molecular dynamic 
simulations (MD) of systems containing 2000 particles with the Gromacs package~\cite{gromacs33}, 
where we kept the temperature constant with a 
Nose-Hoover thermostat~\cite{MP_1984_52_0255,PRA_1985_31_001695} with a relaxation time of 2 ps, 
and the pressure constant at  1 bar with 
a Parrinello-Rahman barostat~\cite{JAP_1981_52_007182} with a relaxation time of 
2 ps. In our MD simulations, we allowed the different box lengths to fluctuate independently.
For the TF alkali halides direct coexistence simulations we used NpT Monte Carlo simulations.

In the free-energy calculations, we used the Einstein Crystal and the Einstein Molecule methods
to compute the free-energy of the solid phase. We performed an initial equilibration run
of the solid in the NpT ensemble of about 10$^5$ Monte Carlo (MC) cycles to obtain the equilibrium density
at the given thermodynamic conditions. We define a MC cycle as a translational trial-move per particle
and a trial-volume change. For the thermodynamic integration in the NVT ensemble we carried out
2x10$^4$ equilibration and 8x10$^4$ production cycles for every value of
$\lambda$ and simulated 20 values of $\lambda$ per thermodynamic state.
We also used thermodynamic integration to compute the free-energy of the liquid phase.
We carried out 8x10$^4$ equilibration and 18x10$^4$ production cycles for every value of $\lambda$ 
and simulated 21 values of $\lambda$ per thermodynamic state. 
Free-energy calculations were performed at 1083~K and 1~bar.
To  obtain the equilibrium densities at these thermodynamic conditions, 
we run  NpT MC simulations of the liquid and solid phases. Once equilibrated,
we used those densities in the free energy calculations.

When performing the thermodynamic integration to compute the free-energy of the liquid phase (route 2), 
we tested the dependence of our results on the choice of the reference system by performing the 
thermodynamic integration to two Lennard-Jones models with different parameters:
the parameters set $LJ1$ used by  Anwar {\em et al.}~\cite{JCP_2003_118_00728} and another
set of parameters denoted as $LJ2$.
Both sets of parameters are represented in Table \ref{LJparam}.
\begin{table}[h!]
\begin{center}
\begin{tabular}{|c|c|c|c||c|c|c|c|}
\hline
\hline
$\epsilon$/k$_B$  [\bf{LJ1}]   & $\sigma$  &  $\rho$*  & T* &  $\epsilon$/k$_B$ [ \bf{LJ2}] & $\sigma$    &   $\rho$*  & T* \\
\hline
 537.01                                        & 2.32                 &  0.3766 & 2.02 &  358.00                                             & 3.00                & 0.8143 & 3.03\\
\hline \hline
\end{tabular}
\end{center}
\caption{\small $LJ1$ and $LJ2$ sets of parameter with units $\epsilon$/k$_B$($[K]$) and $\sigma$([$\AA$]).
$\rho$*=$\rho \sigma^3$ and T*=$k_B$T/$\epsilon$.}
\label{LJparam}
\end{table}

The values of $\rho$* and T* shown in Table \ref{LJparam} are obtained by scaling the density and temperature of
the liquid phase of the  TF/NaCl at 1083~K and 1~bar to LJ reduced units.
 The free-energy should be independent of the choice of the
LJ reference system. Notice that the free-energy of the LJ system given by the 
Nezbeda equation of state (EOS) ~\cite{FPE_100_1994} has a lower error when using LJ2 (at $\rho$*=0.8143 and T*=3.03) rather than at 
$\rho$*=0.3766 and T*=2.02 (when using LJ1) due to the proximity of the LJ fluid critical point.

When using the third route to compute the melting temperature, we 
integrated the Hamiltonian in Eq.~\ref{coupling} from the TF/NaCl to the SD and JC potentials and 
from the TF/NaCl potential to other alkali halides potentials parameterized using TF.
In all cases, we simulated 5 values of $\lambda$ per Hamiltonian Gibbs-Duhem integration.

\section {Results}

Let us start by presenting the results for the melting point of NaCl for the
models and routes considered in this work.

\subsection{Route 1. Liquid/solid direct coexistence}

\begin{figure}[h!]
\centering
\includegraphics[scale=0.35,angle=-0,clip=]{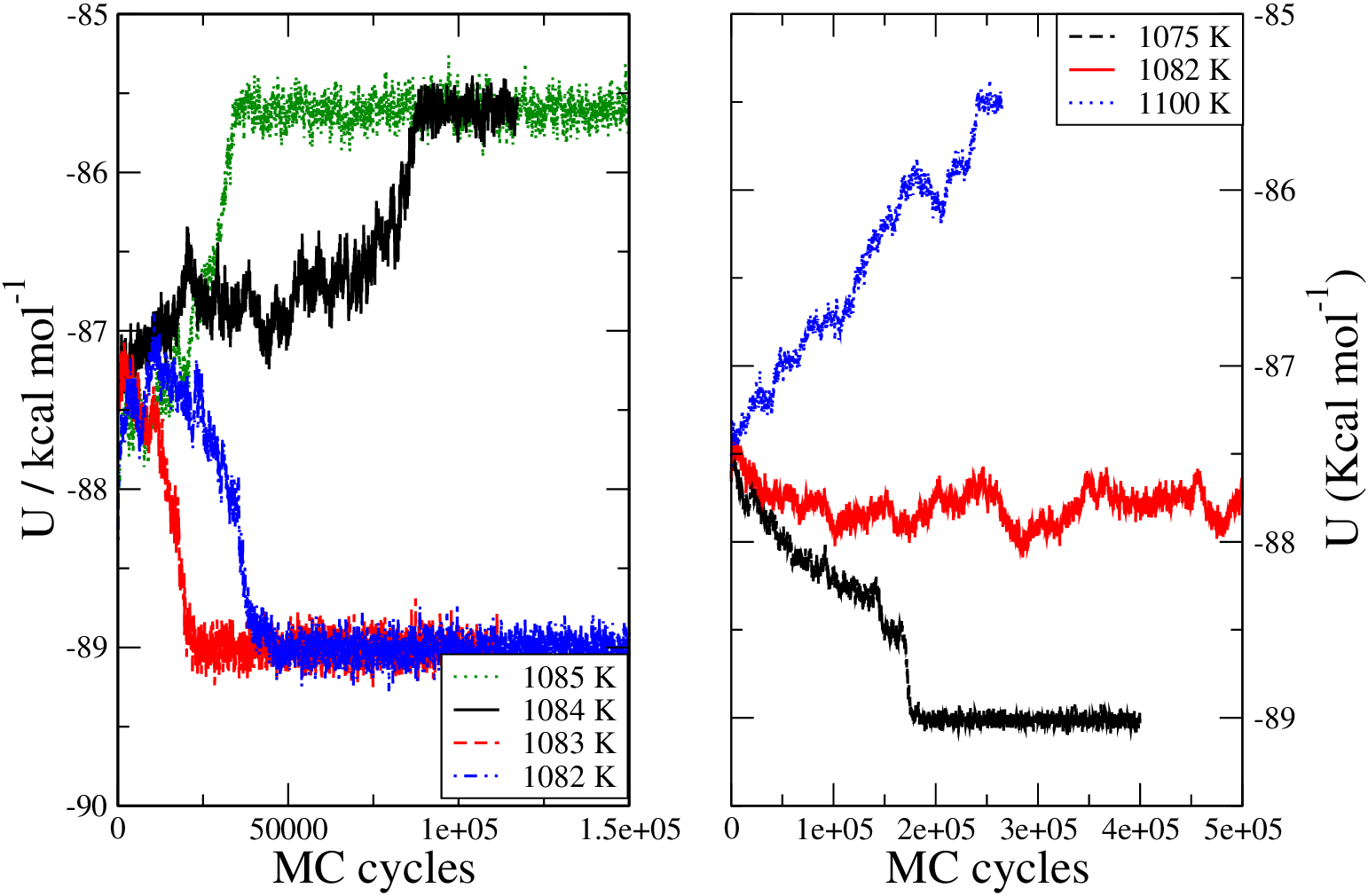}
\caption{\small Total energy (in kcal per ion mol) versus MC cycles for the Tosi-Fumi NaCl 
at T= 1085 K, 1084 K, 1083 K and 1082 K for the 1024 particles system 
(left-hand side) and T= 1100 K, 1082 K and 1075 K  
for the 4000 particles system.}
\label{DC_1024}
\end{figure}

Using the liquid/solid direct coexistence technique~\cite{ladd77,ladd78}
we determined the melting temperature at 1 bar not only of the TF/NaCl but also of the 
other NaCl potentials (the Smith-Dang and Joung-Cheatham). 
After having prepared the equilibrated liquid-solid configuration 
 we performed several anisotropic NpT Monte Carlo simulations (in the case of TF/NaCl) 
for two different system sizes, N=1024 and N=4000, to analyze the finite-size effects.
For SD/NaCl and JC/NaCl we used molecular dynamics simulations with an anisotropic barostat in systems
containing 2000 particles. 
In Fig.~\ref{DC_1024} we plot the time evolution of the total energy of the TF/NaCl system 
equilibrated at 1 bar and at different temperatures for the two system sizes. 

After an equilibration interval of about  10$^4$ MC cycles (where the energy stays constant), we observed that 
when the temperature is below melting the energy decreases until it reaches a plateau with a sudden
change of slope, corresponding to the situation where the liquid has fully crystallized.
When the temperature is above melting, the energy increases until it reaches a plateau 
and stays constant: at this stage, the solid has completely melted.  
The results obtained for the  1024 particles system (left-hand side of Fig.~\ref{DC_1024}) 
show that when T=1083~K the liquid crystallizes, whereas when 
T $>$ 1084~K the solid melts: therefore, the estimated melting
temperature for the 1024 particles system is $T_m$=1083(5)~K. 
The results obtained for the  4000 particles system (right-hand side of Fig.~\ref{DC_1024})
show that when T $<$ 1075~K the liquid crystallizes, whereas when T $>$ 1100~K  
the solid melts. At T=1082K the energy remains approximately constant. From this we conclude 
that for the 4000 particles system $T_m$=1082(13)~K. Thus finite size effects on the melting point of NaCl seems
to be small once the system has at least 1000 particles. 

\begin{figure}[h!]
\centering
\includegraphics[scale=0.35,angle=-0,clip=]{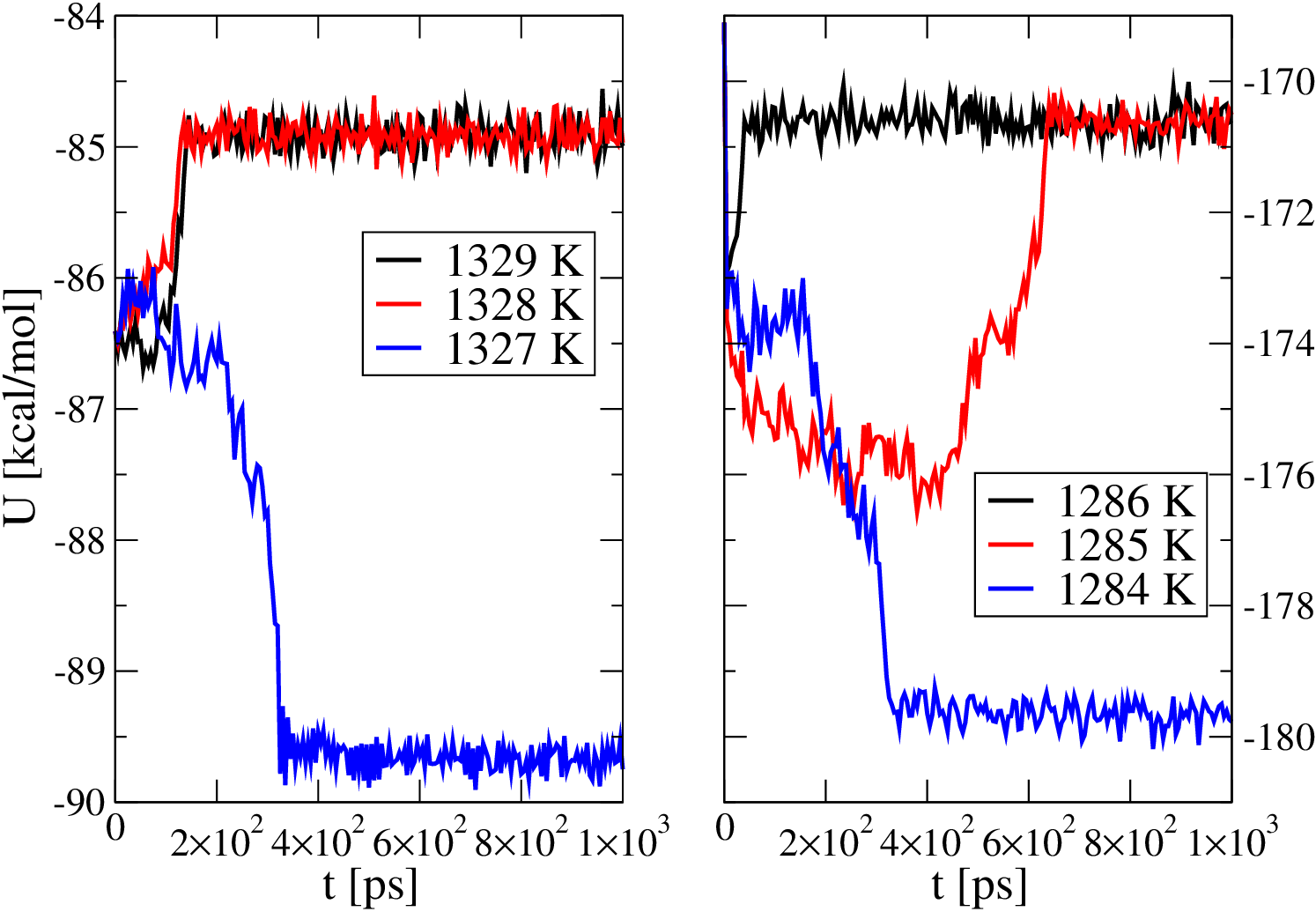}
\caption{\small  Total energy   (in kcal per ion mol) versus time (in picoseconds) for the  NaCl/SD system (left-hand side)
 at T=1329~K, 1328~K and 1327~K, and for the SD/NaCl system (right-hand side) 
at T=  1284~K, 1285~K and 1286~K. Note the different $y$-axis of both plots.}
\label{DC_NaCl_SD-JC}
\end{figure}

Hence, when computing  $T_m$ for the remaining interaction potentials, we chose a large enough
system with 2000 particles. In Fig.~\ref{DC_NaCl_SD-JC},
we present the time evolution of the total energy of the NaCl/SD and NaCl/JC 2000 particles systems, 
at 1 bar and at different temperatures. 
The results obtained for the 2000 particles system show that  the estimated melting temperature is 
 $T_m$= 1327(5)~K for the NaCl/SD and $T_m$=1285(5)~K for the NaCl/JC, respectively.

\begin{figure}[h!]
\centering
\includegraphics[scale=0.35,angle=-0,clip=]{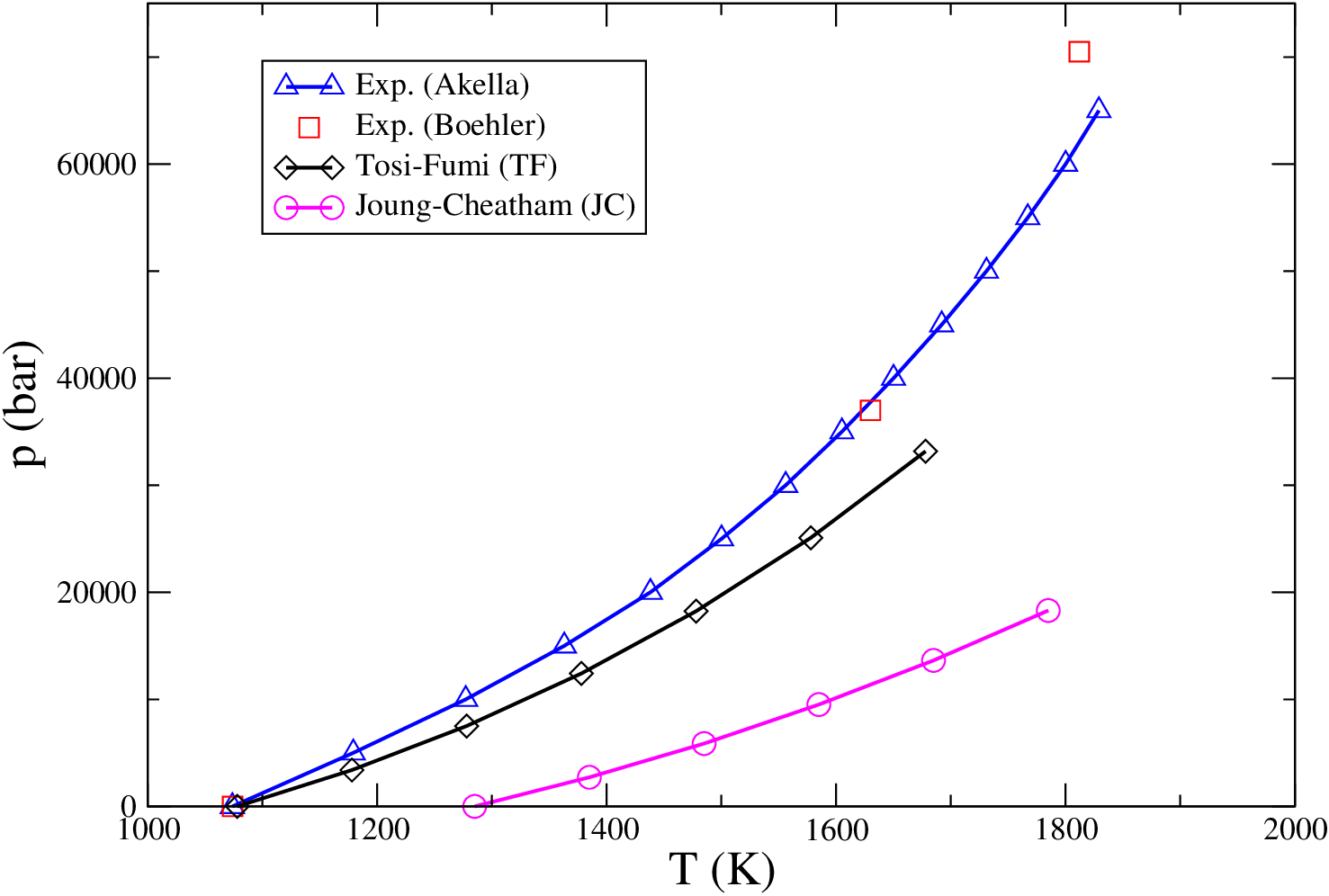}
\caption{\small Melting curve for the TF/NaCl (open diamonds) and JC/NaCl   
(open circles). Experimental results are from Ref.~\cite{PRL_78_4589_1997}
(open squares) and Ref.~\cite{PR_185_1135_1969} (open triangles).}
\label{GD-TF}
\end{figure}

From these results, we can already conclude that the potential   
that gives the melting temperature closest to the experimental one (1074~K) is the Tosi-Fumi.
For this reason the TF/NaCl model is the most suitable for simulations of pure NaCl.
This conclusion is further confirmed when calculating the melting curve for both
TF/NaCl and JC/NaCl potentials.
Using Gibbs Duhem integration~\cite{kofke93}
we calculated the $p-T$ melting curve of the  TF/NaCl and the JC/NaCl,
presented in Fig.~\ref{GD-TF}.

Concerning the results of the TF/NaCl, we observe that   
at low pressure our calculations recover the experimental slope of the melting curve ($\frac{dp}{dT}=30.6(5)$~bar/K),
in good agreement with previous
calculations~\cite{JCP_2003_118_00728,JCP_122_124109_2005},
whereas at higher pressures the slope of the melting line is lower than the experimental one.
On the other hand, concerning the results for the JC/NaCl,
we observe that not only the melting temperature, but also
the slope of the melting curve does not reproduce the experimental one.
The same conclusions can be drawn for the melting enthalpy  ($\Delta h_m$).
The calculated melting enthalpies  for the TF/NaCl, SD/NaCl and JC/NaCl
are 3.36~kcal/mol, 4.8~kcal/mol and 4.6~kcal/mol, respectively, that compared to the experimental value
(3.35~kcal/mol), confirms the better performance of the TF/NaCl model with respect to the
other models.

\subsection{Route 2. Free-energy calculations}

According to the second route, we first computed the Helmholtz free-energy of 
the solid and liquid phase of the TF/NaCl, and then 
 estimated the Gibbs free-energy of each phase ($G$) by adding the  $pV$ term. After that, 
we performed thermodynamic integration of $G$ in the (p,T) plane at constant pressure 
as in Eq.~\ref{gibbs}.
\begin{equation}
\label{gibbs}
\frac{G(T_2,p)}{Nk_BT_2}=\frac{G(T_1,p)}{Nk_BT_1}- \int_{T_1}^{T_2}\frac{H(T)}{Nk_BT^2}dT
\end{equation}
where the enthalpy $H(T) = U(T) + pV(T)$ can be easily obtained in NpT simulations at each temperature. 
The melting point is defined as the state point where 
the two phases have the same chemical potentials ($\mu=G/N$).

We computed the free-energy of the NaCl solid phase at 1083~K and 1~bar with 
Einstein crystal (EC)~\cite{frenkel84}  and Einstein molecule (EM)~\cite{vega_noya} algorithms.
Our results are summarized in Table~\ref{free_sol}, where we present free-energies in $Nk_BT$ units.
To determine the normal
melting point any temperature could have been selected, the choice of 1083~K 
is convenient from
a practical point of view since it is expected to be close to the T$_m$ of the model, so that to reduce 
the contribution of the second term on the right side of the Eq. \ref{gibbs}.
\begin{table}[h!]
\centering
\resizebox{8.5cm}{!}{
\begin{tabular}{|c|c|c|c|c|c|c|c|c|}
      \hline
      \hline
 & $T$[K] & N  & $\rho$ (N/\AA$^3$) & $\Lambda_E [k_B T \AA^{-2}]$ & A$_0$[Nk$_B$T] & $\Delta$ A$_1$[Nk$_B$T] &  $\Delta$ A$_2$[Nk$_B$T] & A$^{NaCl}_{sol}$[Nk$_B$T]   \\
\hline
\hline
TF/EC & 1083 & 1000 & 0.03856 & 500 & 7.583 & -42.73 & -6.33 & -41.481(9) \\
TF/EM & 1083 & 1000 & 0.03856 & 500 & 7.594 & -42.73 & -6.34 & -41.477(9) \\
\hline
\hline
\end{tabular}
}
\caption{\small \label{free_sol} Free energy from the Einstein Crystal/Einstein Molecule 
for the TF/NaCl solid phase at 1 bar and different temperatures ($T$) and system size
($N$ being the total number of particles). $\rho$ the number density ($N/V$), $\Lambda_E$
the spring constant, $A_0$, $\Delta A_1$, $\Delta A_2$ are the terms in Eq.~\ref{einstein} and
the free energy of the solid $A^{NaCl}_{sol}$ is represented in the last column and corresponds
to Eq.\ref{einstein}.}
\end{table}

In Table~\ref{free_sol}, we observe a perfect agreement  between 
the free-energy  computed with the Einstein Crystal and the one compute with the Einstein Molecule 
at the same pressure, temperature and system size (1000 particles). 

Having computed the chemical potential of the solid phase, we calculated the Helmholtz 
free-energy of the liquid phase using two Lennard-Jones reference systems $LJ1$ and $LJ2$
in order to study the uncertainties associated to the choice of the reference system. 
Results are summarized in Table~\ref{free_liq}.
\begin{table}[h!]     
\centering
\resizebox{8.5cm}{!}{
       \begin{tabular}{|c|c|c|c|c|c|}
      \hline
      \hline
 & $\Delta$ A$_{liq}^{LJ}$[Nk$_B$T] & A$_{liq}^{LJ,id}$[Nk$_B$T] & A$_{liq}^{LJ,res}$[Nk$_B$T]  &  A$_{liq}^{NaCl}$[Nk$_B$T] \\
\hline
\hline
 TF/LJ1 &  35.95 & -5.19 & -0.327 & -41.47(2)  \\
 \hline
 TF/LJ2 &  37.12 & -5.19 & 0.837 & -41.48(2)  \\
\hline
\hline
\end{tabular}
}
\caption{\small   \label{free_liq} TF/NaCl free energy of the liquid phase (A$_{liq}^{NaCl}$)
as in Eq.~\ref{freeliq} at 1~bar and T=1083~K. The results presented in the table refer to a
system with $N=1000$ particles and number density of $\rho= 0.03016$~\AA$^{-3}$. $\Delta$A$^{LJ}$ is
the integral in Eq.~\ref{freeliq}. $\frac{A_{liq}^{LJ,id}}{Nk_BT}= ln(\rho/2)-1.00$. 
A$_{liq}^{LJ,res}$ is obtained from the EOS of Ref.~\cite{FPE_100_1994}.} 
\label{resultsfreeliq}
\end{table}

\begin{figure}[h!]
\centering
\includegraphics[scale=0.35,angle=-0,clip=]{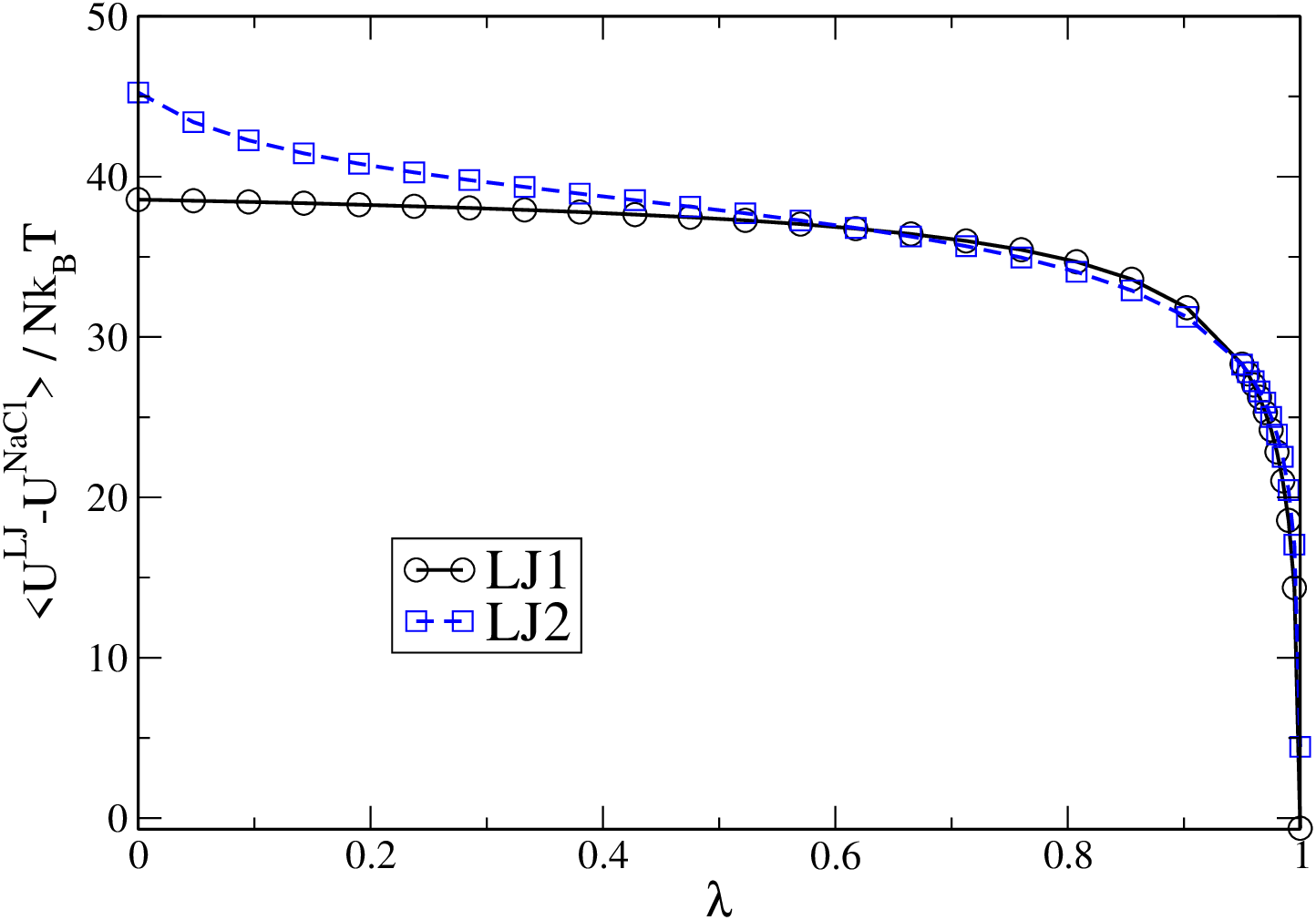}
\caption{\small $\langle U^{LJ} - U^{NaCl} \rangle_{N,V,T,\lambda}$ from Eq.~\ref{freeliq} at 1083 K and 1 bar. The two curves correspond to 
a different set of Lennard-Jones parameters for the reference system: $LJ1$ (open circles) and $LJ2$ (open squares).}
\label{integral}
\end{figure}

As shown in  Table \ref{resultsfreeliq}, the value of the integral ($\Delta$A$_{liq}^{LJ}$)
depends on the choice of the LJ parameters of the reference LJ system. In Fig.~\ref{integral}
the integrand of Eq. \ref{freeliq} is shown
when carrying out the thermodynamic integration from the liquid TF/NaCl to both LJ reference
systems $LJ1$ and $LJ2$.  It is relevant to stress that, due to 
the abrupt change of $\langle U^{LJ} - U^{NaCl} \rangle_{N,V,T,\lambda}$ 
for values of $\lambda$ close to 1, many points were used in the integration between $\lambda=0.95$ and $\lambda=1.00$.

To calculate the residual free energy of the reference LJ fluid at our thermodynamic conditions,  
we used the equation of state (EOS) for the LJ system proposed by Nezbeda {\em et al.}~\cite{FPE_100_1994,sklogwiki}.
Independently on the chosen reference system (whether $LJ1$ or $LJ2$),
the two free-energies A$_{liq}^{NaCl}$[Nk$_B$T] coincide (last column in Table \ref{resultsfreeliq}).

After having computed the Helmholtz free-energy, we 
 estimated the Gibbs free-energy ($G$) by adding the  $pV$ term and  
performed thermodynamic integration of $G$
at constant pressure 
(see Eq.~\ref{gibbs}) where we used NpT MC simulations to compute enthalpy  $H(T)$  and 
to estimate the equation of state of the liquid and solid phases
(where a typical MC run consists of 3x10$^4$ equilibration and 7x10$^4$ production cycles).
The chemical potential of one phase is $\mu = G/N$ and the melting
temperature is given by the point at which the two phases have the same chemical potentials. Our results 
for the TF/NaCl are presented in Table~\ref{melting_route1}.
\begin{table}[h!]
\centering
 \begin{tabular}{|c|c||c|c|}
  \hline
  \hline
 & T$_m$[K]   & &  T$_m$[K]\\
\hline
\hline
 TF/LJ1 & 1083(3)  & TF/LJ2 & 1084(3) \\
\hline
\hline
\end{tabular}
\caption{\small \label{melting_route1} Melting temperature at 1 bar of the TF/NaCl model. The results presented in the table refer to a system with $N=1000$ particles.
The parameters for the interaction potentials of the reference systems ($LJ1$ and $LJ2$) are
given in Table \ref{LJparam}.}
\end{table}

Both results are in perfect agreement with the T$_m$ calculated by direct coexistence.
Other EOS could be used to calculate the residual free energy of the reference LJ fluid,
such as the one proposed by Johnson {\em et al.}~\cite{JPC_1994_98_06413}.  When the LJ residual contribution
is taken from Johnson {\em et al.}, the melting temperature turns out to be about 6~K higher
than when it is taken from Nezbeda {\em et al.} However, it is likely that the EOS proposed by Nezbeda is
slightly more accurate than that by Johnson {\em et al.}~\cite{FPE_100_1994}. In any case, the differences are small.

\subsection{Route 3. Hamiltonian Gibbs-Duhem integration}

\begin{figure}[h!]
\centering
\includegraphics[scale=0.35,angle=-0,clip=]{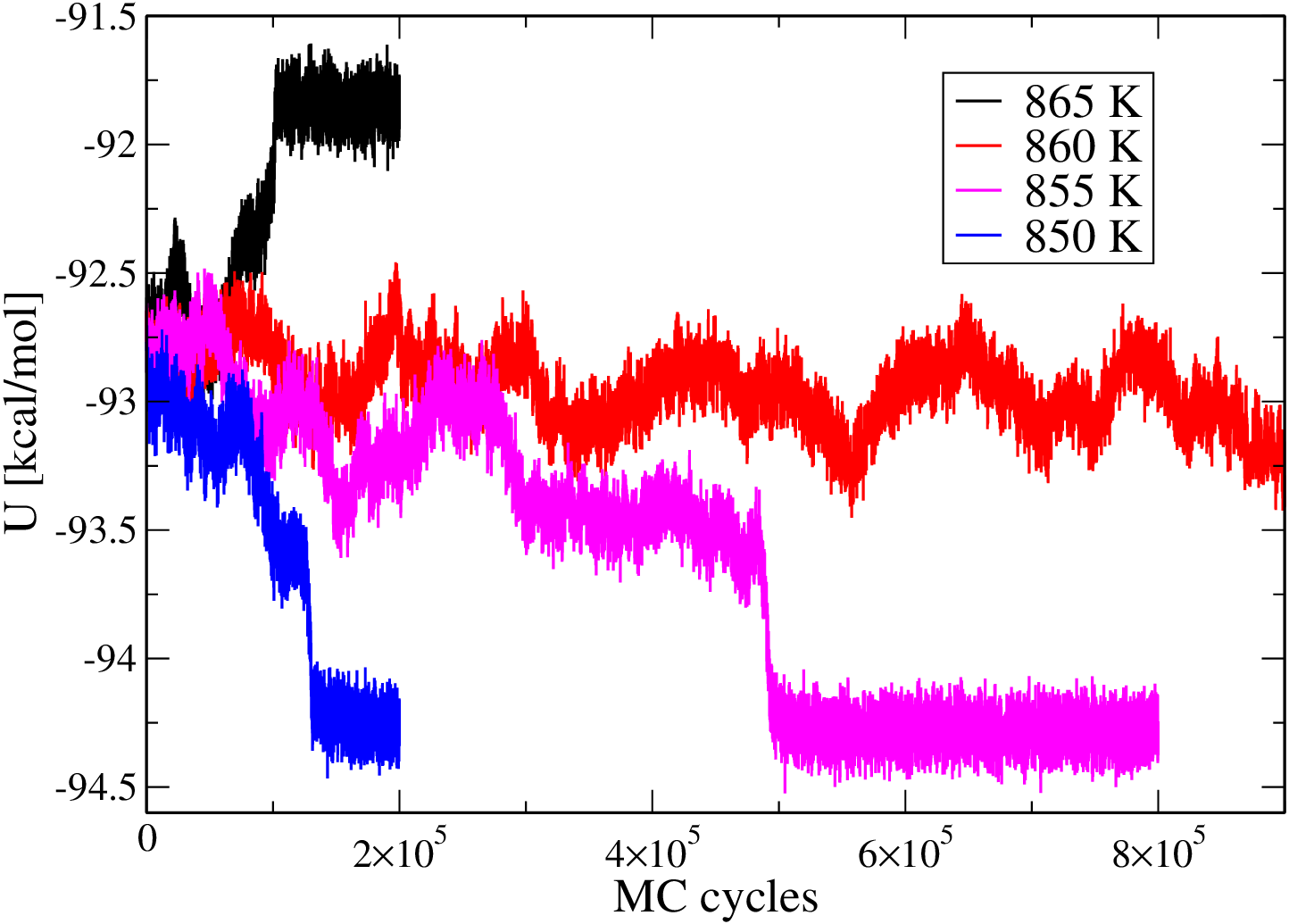}
\caption{\small  Total energy versus MC cycles for TF/KF  
at T =  865 K, 860 K, 855 K and 850 K (from top to bottom).}
\label{DC_KF}
\end{figure}

The third route we followed to compute the melting temperature of several alkali halides at 1 bar 
is by means of the Hamiltonian Gibbs-Duhem thermodynamic integration.
We used the TF/NaCl as the reference Hamiltonian ($U_A$ in Eq.~\ref{coupling}) and  
integrated the generalized Clapeyron equation (Eq.~\ref{ham_GD_int2}) using a 
1000-particle system. 
From the $\lambda=1$ point of the Hamiltonian Gibbs-Duhem integration, we could compute the melting entropy  
$\Delta S_m$ (being $\Delta S_m = N \Delta s_m$, see Eq.~\ref{meltentropy}) and the melting
enthalpy  ($\Delta H_m = N \Delta h_m$) at coexistence.
In the Table \ref{melting_points} we present our results for the melting temperature, 
$\Delta S_m$, reporting also the experimental values of both for each alkali halide, and for 
$\Delta H_m$.
We checked our Hamiltonian Gibbs Duhem calculations by computing 
the melting temperature of TF/KF at 1~bar by means of liquid/solid direct coexistence.
Figure~\ref{DC_KF} represents the time evolution of the total energy of the TF/KF 
2000-particle systems, equilibrated at 1 bar and at different temperatures. 
From these results we estimate T$_m$ to be 860(5) K, that corroborates the result obtained 
with the Hamiltonian Gibbs Duhem integration in Table~\ref{melting_points} (i.e. 859(15)K).

\begin{table}[h!]
  \begin{tabular}{|c|c|c||c|c|c|c|}
  \hline
  \hline
Alkali halide & T$_m$ &T$^{exp}_m$ & $\Delta S_{m}$ & $\Delta$S$_{m}^{exp}$ &  $\Delta H_{m}$ & $\Delta$H$_{m}^{exp}$ \\
\hline
\hline
{\bf NaCl} & {\bf 1082(13)}  & {\bf 1074} & {\bf 3.10} & {\bf 3.12} & {\bf 3.36} & {\bf 3.35} \\
KF         & 859(15)  & 1131 & 2.77 & 2.98 & 2.39 & 3.37 \\
{\bf KBr}  & {\bf 1043(15)} & {\bf 1003} & {\bf 3.14} & {\bf 3.03} & {\bf 3.29} & {\bf 3.04} \\
{\bf KCl}  & {\bf 1039(15)} & {\bf 1049} & {\bf 3.12} & {\bf 3.04} & {\bf 3.25} & {\bf 3.19} \\
RbBr       & 1047(15) & 955  & 3.34 & 2.88 & 3.51 & 2.75 \\
RbCl       & 1092(15) & 988  & 3.18 & 2.88 & 3.48 & 2.85 \\
RbF        & 996(15)  & 1048 & 3.03 & 2.88 & 3.03 & 3.02 \\
NaF        & 611(15)  & 1266 & 2.10 & 3.09 & 1.29 & 3.91 \\
{\bf NaBr} & {\bf 1022(15)} & {\bf 1028} & {\bf 3.10} & {\bf 3.06} & {\bf 3.18} & {\bf 3.70} \\
LiCl       & 780(15)  & 887  & 2.46 & 2.70 & 1.93 & 2.39 \\
 LiF  &  1010(15) &  1118 &  2.93 &  2.88 &  2.97 &  3.22 \\
\hline
\hline
\end{tabular}
\caption{\small  \label{melting_points} Melting temperatures, $\Delta S_m$ and $\Delta H_m$ at 1 bar for
several alkali halides simulated using the TF interaction potential with units T$_m$($[K]$),
$\Delta S_m$($[cal/(K mol)]$) and $\Delta H_m$($[kcal/mol]$).
The experimental values of $\Delta S_m$ are from Ref.~\cite{Ubbelohde,coulomliquids,wunderlich}.
The bold fonts represent alkali halides for which there are good agreement
with the experimental melting temperatures.}
\end{table}

While the Tosi-Fumi potential predicts the NaCl melting temperature in good agreement with its experimental value, 
the results for the other alkali halides are not as accurate.
The predictions for KCl and NaBr are reasonable (differing only 
about 15~K and 10~K from the experiments). Whereas for alkali
halides involving Rb or F ions (such as KF, RbBr, RbCl, RbF, LiF and NaF)
the calculated values strongly differ from the experimental ones  
(with deviations of up to 100~K). From this  we conclude that the TF model potential, being  non-polarizable,  fails when describing 
alkali halides that involve big cations (such as Rb$^+$) and small anions (such as F$^-$).
In general, when T$_m$ of a given TF/alkali halide is lower/higher than the experimental value, also
$\Delta H_m$ is smaller/larger  than the experiments, so that 
the predicted $\Delta S_m$ is in reasonable agreement with the experimental value for most
of the studied alkali halides.

\section{Discussions}

We now compare our data for the melting temperature of NaCl at 1~bar with other values taken from the 
literature (see Table \ref{NaClmodels}).

\begin{table}[h!]
\begin{tabular}{|c|c|c|c|c|}
\hline
\hline
                                                & T$_m$[K] & N & $\rho_{liq}$ & $\rho_{sol}$  \\
\hline
\hline
TF/Anwar~\cite{JCP_2003_118_00728}        & 1064(14) & 512 &  --     & --         \\
TF/Anwar*~\cite{JCP_2003_118_00728}        & 1084(14) & 512 &  --     & --         \\
TF/Zykova-Timan~\cite{zykova3}        & 1066(20) & 5760 &  --     & --         \\
TF/Mastny~\cite{JCP_122_124109_2005}    & 1050(3) & 512 &  --     & --         \\
TF/Eike~\cite{brennecke}                & 1089(8) & 512 &  --     & --         \\
TF/An~\cite{JCP_125_154510_2006}         & 1063(13) & 4096 &  --     & --          \\
TF/this work                            & 1083(5)  & 1024 &  1.465 & 1.876       \\
TF/this work                            & 1082(13)  & 4000 &  1.465 & 1.876       \\
SD/this work                            & 1327(10) & 2000 &  1.216 & 1.668        \\
JC/this work                            & 1286(10) & 2000 &  1.283 & 1.746        \\
Experiments/Janz~\cite{janz_molten_salts} & 1074    & - & 1.54 & 1.98           \\ 
\hline
\hline
\end{tabular}
\caption{\small \label{NaClmodels} Melting temperature of the NaCl computed using TF, SD and JC 
interaction potentials at 1 bar. N represents de number of ions of each study.
$\rho_{liq}$ and $\rho_{sol}$ are the liquid and solid densities in $g cm^{-3}$, respectively. The * represents
the recalculated T$_m$ at 1~bar using the values given by Anwar (p$_{coex}$=-300~bar at 1074~K and
$\frac{dp}{dT}=30~{\rm bar}~K^{-1}$).}
\end{table}

Concerning the calculations of the melting temperature with the Tosi-Fumi, 
we observe that $T_m$ calculated in this work is in good agreement (within the error bar) 
with the one reported by Anwar {\em et al.}~\cite{JCP_2003_118_00728}.
They calculated the coexistence pressure at 1074~K and 
$-300$~bar. Then, they evaluated the slope of the coexistence curve, $\frac{dp}{dT}=30~{\rm bar}~K^{-1}$,
and recalculated the melting temperature at 1~bar obtaining 1064(14)~K (nonetheless, using their values,
we have obtained the NaCl melting temperature at 1084(14)~K, in perfect agreement with our results).
Zykova-Timan {\em et al.}~\cite{zykova3} computed the melting temperature via 
 liquid/solid direct coexistence and obtained a value for T$_m$ in good agreement with the one obtained in this work (taking into account that in their NVE runs they obtained the melting temperature for a pressure of about
+- 500 bar and since the slope of the melting curve is of about 30K/bar the value at $p=1bar$ could be modified by
up to 17K ). 
In general, the agreement with data in the literature is satisfactory.
The TF  slightly underestimates the experimental values of the  coexistence densities,
whereas both  SD and JC underestimate them considerably, in agreement with the results of  Alejandre and Hansen for the SD model \cite{PRE_76_061505_2007}.

When computing the melting temperature of the TF/NaCl model potential via liquid/solid direct coexistence
and free-energy calculations we have studied 
finite-size effects and observed that the value of the melting temperature did not change much for systems
having more than 1000 particles. Thus we take T$_m$=1082(13)~K (i.e the melting point of the 4000 particle
system ) as our estimate of the melting point of the TF/NaCl model potential in the thermodynamic limit ( where the error bar now includes the possible contribution to the error
of the extrapolation to infinite size). 
When calculating  T$_m$ via Hamiltonian Gibbs-Duhem integration, we have used the value of T$_m$=1082(13)~K 
as the initial point for the Hamiltonian Gibbs Duhem integration at coexistence. 

Concerning the calculation of the melting temperature with the 
Smith-Dang and Joung-Cheatham potentials, we computed these values via two independent routes
(liquid/solid direct coexistence and Hamiltonian Gibbs-Duhem integration) 
and obtained the same results. The T$_m$ of both models is reported here.
From our calculations we conclude that these models overestimate 
the melting temperature of NaCl, being about 200-250~K higher than the experimental value of 1074~K.
Therefore, it is clear that 
the melting temperature that most resembles the experimental one at 1 bar is the one calculated using 
the TF/NaCl. Although JC/NaCl or SD/NaCl would work for NaCl solutions in water,
they seem to be unsuitable for simulations of pure NaCl.
 
\section{Conclusions}

In this manuscript, we have computed the melting temperature at 1 bar for different NaCl-type alkali halides.
When computing  T$_m$ for the Tosi Fumi NaCl interaction potential, we have followed three independent routes:
1) liquid/solid direct coexistence, 2) free-energy calculations and 
3) Hamiltonian Gibbs-Duhem integration. For the Smith Dang /NaCl and Joung Cheatham NaCl,
we have used the first and third route, whereas for other Tosi Fumi alkali halides we have applied only the third route.

The results obtained for the T$_m$ of TF/NaCl are in good agreement with other numerical \cite{JCP_2003_118_00728,SS_566_794_2004}
and with experimental results \cite{janz_molten_salts}, giving  T$_m$= 1082(13)~K at 1 bar. 
When computing  T$_m$ for the SD/NaCl and JC/NaCl, we find a perfect agreement between the calculations obtained via
liquid/solid direct coexistence and Hamiltonian Gibbs-Duhem integration.
However, both models overestimate the melting temperature of NaCl by more than 200~K.
We have also determined the melting curve for the Tosi-Fumi and Joung-Cheatham models and found that 
the Tosi-Fumi correctly predicts the behavior of the curve ($\frac{dp}{dT}$) at low pressures, but does not  
capture the experimental behavior when the pressure increases. 
Therefore, we conclude that the SD/JC models are unable to reproduce the properties of pure NaCl.

We have also computed the melting temperature of other alkali halides using the Tosi-Fumi 
interaction potential and observed that this model gives good predictions for NaCl, NaBr, KCl and KBr, whereas 
for the other alkali halides the predictions are not as good, especially when it concerns Rb$^+$ and F$^-$ ions. 
The reason for this probably being that the Tosi-Fumi interaction potential is not polarizable and cannot 
capture the highly polarizable character of these ions.
Neglecting the polarization terms causes an incorrect description of these salts. 
 In the past polarization effects
were normally incorporated into the simulations of ionic systems via the shell
model \cite{AP_23_247_1976}. More elaborate models developed recently employ potentials which
include the polarization effects using either multipole expansions \cite{JPCM_5_2687_1992} or the
distortable ion model \cite{JCP_119_9673_2003}. 
The TF/alkali halide potentials have a serious transferability problem: the same ions present
different potential parameters depending on the alkali halide (i.e. the Na-Na interaction
is different in NaCl and NaF).
In the case of LJ-like models, it seems that it  is not possible to describe NaCl accurately with
a model consisting of point charges and a Lennard-Jones interaction site for each ion
where Lorentz-Berthelot rules are used to describe the interaction between cations and anions.
The same conclusion was obtained by Cavallari {\em et al.} \cite{MP_102_959_2004} who showed how the ion-ion interaction
reconstructed from standard LB rules fail to correctly account for the structure of the concentrated solutions
model with {\em ab-initio} calculations.
An interesting possibility would be incorporating deviations to the Lorentz-Berthelot combining
rules to obtain the crossed interactions between the ions or adjusting these interactions as in the TF model,
although probably the use of a Van der Waals r$^{-6}$ term would not capture the underlying physics.
Thus, it seems there is still room for improvement in the area of alkali-halides salt models.

\section{acknowledgments}
This work was funded by grants FIS2010-16159.
from the DGI (Spain), MODELICO-P2009/ESP/1691 from the CAM,
and 910570 from the UCM. J. L. Aragones would like to thank the MEC by the award of a pre-doctoral
grant. C.Valeriani acknowledgments financial support from a Juan de la Cierva 
Fellowship and from  a PCIG-GA-2011-303941 Marie Curie Integration Grant. E. Sanz
acknowledgments financial support from a Ramon y Cajal Fellowship.
We would like to thank the referees for their helpful comments.

\bibliographystyle{./apsrev}

\end{document}